\documentclass[aps,prd,reprint,twocolumn,nofootinbib]{revtex4-2}
\usepackage{amsmath,amssymb,amsfonts,mathrsfs,graphicx}
\usepackage{hyperref}
\usepackage[dvipsnames]{xcolor}     

\begin{document}

\title{The Cosmological Constant from a Quantum Gravitational $\theta$-Vacua \\and the Gravitational Hall Effect  }

\author{Stephon Alexander}
\author{Heliudson Bernardo}
\author{Aaron Hui}
\affiliation{Department of Physics, Brown Theoretical Physics Center, Brown University, Providence, RI 02912, U.S.A}

\begin{abstract}

We provide a new perspective on the cosmological constant by exploring the background-independent Wheeler-DeWitt quantization of general relativity. The Chern-Simons-Kodama state of quantum gravity, a generalization of the Hartle-Hawking and Vilenkin states, has a striking structural similarity to the topological field theory of the quantum Hall effect. As a result, we study the gravitational topological $\theta$-sectors in analogy to Yang-Mills theory. We find that the cosmological constant $\Lambda$ is intimately linked to the $\theta$-parameter by $\theta=12\pi^2/(\Lambda \ell^2_{\rm Pl}) \mod 2\pi$ due to the fact that Chern-Simons-Kodama state must live in a particular $\theta$-sector. This result is shown in the canonical, non-perturbative formalism. Furthermore, we explain how the physics of the Hamiltonian constraint is analogous to the quantum Hall effect, with the cosmological constant playing the role of a quantum gravitational Hall resistivity. These relations suggest that $\Lambda$ is topologically protected against perturbative graviton loop corrections, analogous to the robustness of quantized Hall conductance against disorder in a metal.

\end{abstract}

\maketitle

\section{Introduction}

The quantum Hall effect (QHE) is a surprising phenomenon in which the Hall conductivity of a quantum many-body system becomes exactly quantized in terms of two fundamental constants of nature \cite{Hall1, vonKlitzing:1980pdk, Laughlin:1981jd, Tsui:1982yy, Laughlin:1983fy,Stormer:1999zz}. The quantized Hall conductance is observed to be extraordinarily precise, up to one part per billion \cite{Janssen_2012}, and is given by:
\begin{equation}
\sigma_H = \frac{e^2}{h} \nu
\end{equation}
where \(\sigma_H\) is the Hall conductance, \(e\) is the elementary charge, \(h\) is Planck's constant, and \(\nu\) is the filling factor.

This quantization is profound. For most electronic systems, quantization should not occur; generic scattering from, e.g., disorder or impurities disrupts electronic transport and quantitatively affects the conductivity. However, in the QHE, the quantized Hall conductance remains invariant despite significant levels of disorder. The reason why the conductance is robust is because of topological quantum effects: the topological properties of the electronic wavefunctions are immune to local perturbations. As long as the system remains in the same topological phase (i.e., no phase transitions occur), disorder cannot change the quantized value of conductance \cite{Hofstadter:1976zz,Thouless:1982zz, Bellissard_1994}.

The effective theory of the Hall current in the QHE is a Chern-Simons theory \cite{Witten:1988hf}, a topological field theory that leads to a quantized Hall conductance (see for instance \cite{PhysRevLett.62.82, PhysRevB.42.8145,Balachandran:1991nw,Frohlich:1995mec, Susskind:2001fb}). Chern-Simons theory also appears in the canonical approach to quantum general relativity (GR) within the solution to the quantum Hamiltonian constraint in the presence of a cosmological constant. In fact, a proposed ground state of quantum gravity, the Chern-Simons-Kodama (CSK) state, is written in terms of a non-Abelian Chern-Simons functional \cite{Kodama}. The CSK state has long been an object of interest in canonical quantum gravity \cite{Kodama, Smolin:2002sz, Freidel:2003pu, Randono:2006rt, Randono:2006ru, Alexander:2008yg, Magueijo:2020ugp, Alexander:2022ocp} (see also \cite{Witten:2003mb} for early open issues). This suggests that quantum gravity, by way of the CSK state, may also exhibit non-trivial features as a result of topology.

In this letter, we use the Wheeler-DeWitt (WdW) formalism \cite{DeWitt, Wheeler:1968iap} and show that the cosmological constant $\Lambda$ in general relativity is topologically protected, similar to the Hall conductance. By analogy to the topological $\theta$-sectors in Yang-Mills (YM) theory, proper treatment of quantum gravity also requires a prescription of a $\theta$-sector. As shown in this work, consistency of the CSK state with the choice of $\theta$-sector leads to the constraint
\begin{align}\label{eq:main_result}
    \theta = \frac{12\pi^2}{\Lambda \ell_{\rm Pl}^2} \mod 2\pi
\end{align}
where $\ell_{\rm Pl}$ is the reduced Planck length. Surprisingly, the topological parameter $\theta$ is intimately linked to the cosmological constant; the ``superselection'' choice of $\theta$ ``quantizes'' the allowed values of $1/\Lambda$.

Remarkably, we explain how the CSK state suggests a gravitational analogue of the Hall current. Along those lines, in Sec.~\ref{sec: quantum gravitational Hall}, we discuss how the physics of the QHE and canonical quantum gravity in the self-dual variables are analogous to each other. From this perspective, the cosmological constant plays a role analogous to the filling factor \(\nu\) in the quantum Hall effect.

We emphasize that we utilize a non-perturbative and topological approach to investigate quantum gravity and to compute cosmological observables. This approach not only allows us to make exact statements (e.g. the existence of a $\theta$-sectors), but also makes the calculation of some non-trivial observables such as the probability current to be more straightforward. Based on our results, we also highlight the need for further investigation into the implications of these topological structures in gravitational physics.

\section{Large Gauge Transformations and the Euclidean CSK State}\label{sec:QG_theta_sector_canonical}

By consistency of a gauge theory, we must ensure that all physical observables are gauge-invariant. This is partially addressed by the Gauss constraint $\hat{\mathcal{G}}_i \Psi[A] = 0$, which guarantees that the wave function is invariant under small gauge transformations, i.e. $e^{i\hat{\mathcal{G}_i}} \Psi[A] = \Psi[A]$. However, one can also have homotopically nontrivial gauge transformations, i.e. large gauge transformations \cite{Jackiw:1983nv}, for which the Gauss constraint has nothing to say. As we discuss below, the non-trivial topology associated with the gauge transformations gives rise to $\theta$-sectors.

\subsection{YM and topological $\theta$-sectors}

We first review the topological structure of gauge transformations and YM theory to draw analogies to the gravitational case, following Ref.~\cite{Ashtekar:1988sw}. Consider local gauge transformations $\Sigma_3 \rightarrow \text{SU}(2)$ that are asymptotically identity, i.e. $g(r) \to 1$, as $r\to \infty$. For $\Sigma_3 = \text{S}^3$, these maps can be classified by the winding number $w(g) \in \mathbb{Z}$ \footnote{Even when $\Sigma_3 = \mathbb{R}^3$, all such maps can be extended to $\text{S}^3$, which is the one-point compactification of $\mathbb{R}^3$. More explicitly, these maps are classified by $\pi_3(\text{SU}(2)) = \mathbb{Z}$.}, defined by
\begin{equation}
    w(g) = \frac{1}{24\pi^2}\int d^3x \epsilon^{abc}\text{tr}(g^{-1}\partial_a g) (g^{-1}\partial_b g) (g^{-1}\partial_c g).
\end{equation}
The elements with $w(g) \neq 0$ are called large gauge transformations. A nonzero winding number implies that \(g(x)\) ``wraps nontrivially'' around SU(2), homotopically distinguishing large gauge transformations from small ones. As an example, a standard element for \(g(x)\) with winding number \(w = 1\) is
\begin{equation}\label{eq:large_gauge}
g(x) = e^{i f(r) \hat{x}^i \sigma^i},
\end{equation}
where \( \sigma^i \) are the Pauli matrices, \( \hat{x}^i = x^i /r \) is a unit radial vector where $r = |x|$, and \( f(r) \) is a function that interpolates smoothly from \( f(0) = 0 \) at the origin to \( f(\infty) = \pi \) at infinity. This ensures that the gauge transformation covers SU\((2) \) exactly once, yielding \( w(g) = 1 \). Transformations with $w(g) = n$ can be obtained by composing this $g(x)$ with itself $n$ times.

To properly define YM theory, one must specify how wavefunctions $\Psi_{\rm YM}$ transform under large gauge transformations. Due to the condition that physical observables are invariant under large gauge transformations, the wavefunction must be ``quasi-periodic'' (i.e. equivalent up to phase) under these transformations. This means we must make a choice of $\theta$ where \cite{Jackiw:1976pf, Callan:1977gz}
\begin{equation}\label{eq:psi_YM}
    \Psi_{\rm YM}^\theta (A^g) = e^{i\theta w(g)}\Psi_{\rm YM}^\theta(A).
\end{equation}
where $A^g$ is the gauge-transformed connection under the action of $g$. In particular, there is a $\theta$-ambiguity about how to define the action of large gauge transformations; the choice of $\theta \in \mathbb{R}/(2\pi\mathbb{Z})$ defines (superselects) the $\theta$-sectors. The ambiguity is parameterized by a single parameter since the large gauge transformations are generated by a single element, i.e. are labeled with homotopy classification $\mathbb{Z}$. As a remark, we can equivalently think of $\theta$-sectors as arising from the topology of the space of gauge-equivalent connections (i.e. connections $A$ that are identified under all gauge transformations) by thinking of $\Psi$ as a multi-valued wavefunction \cite{Klimek-Chudy:1984wcz}.

\subsection{Gravitational $\theta$-sectors}

Before we consider the analogous situation in gravity, let us write the canonical formulation of general relativity as close as possible to a gauge theory. This is accomplished by employing the so-called Ashtekar variables \cite{Ashtekar:1986, Ashtekar:1987}. In the Ashtekar approach, the Hamiltonian formulation of GR can be written in terms of an SU(2) connection $A_a^i$, defined on spacelike hypersurfaces that foliate the spacetime in a 3+1 decomposition. The induced metric on these hypersurfaces is written in terms of the triads $E^a_i$ as $q^{ab} = E^a_i E^b_j \delta^{ij}$, where the index $a$ runs from 1 to 3. The connection and triad satisfy the Poisson bracket relation
\begin{equation}
\{A^i_a(x), \tilde{E}^b_j(y)\} = 8\pi G\beta \delta^b_a \delta^i_j \delta^{(3)}(x-y),
\end{equation}
where $\beta =i$ ($\beta=-1$) in the Lorentzian (Euclidean) case. Due to spacetime reparameterization invariance, the total Hamiltonian of GR is a linear combination of constraints. The piece associated to time-reparameterization invariance is proportional to 
\begin{equation}
\mathcal{\tilde{H}} = \frac{\beta^2}{16\pi G \sqrt{|\tilde{E}}|}\epsilon^{ijk} \tilde{E}^a_i \tilde{E}^b_j \left(F_{ab}^k + \frac{\Lambda}{3} \epsilon_{abc} \tilde{E}^c_k \right),
\end{equation}
where $\Lambda$ is the cosmological constant, $F^k_{ab}$ is the curvature of $A_a^k$, and $\mathcal{\tilde{H}} \approx 0$ is called the Hamiltonian constraint. At quantum level, we should have $\hat{\tilde{\mathcal{H}}} \Psi[A] = 0$, which is the Wheeler-DeWitt equation in the Ashtekar formulation. 

As we discuss in the Appendix \ref{sec:review}, an exact solution to all the GR constraints for $\Lambda \neq 0$ is given by the so-called Chern-Simons-Kodama (CSK) state \cite{Kodama}. In the Euclidean case, the CSK state is given by
\begin{equation}\label{eq:Euclidean_CSK_with_Lambda}
    \Psi_{\rm CSK}[A] = \mathcal{N} e^{\frac{3i}{2\Lambda \ell^2_{\rm Pl}}{\rm CS}[A]},
\end{equation}
where ${\rm CS}[A]$ is the Chern-Simons function of $A^i_a$,
\begin{equation}
    {\rm CS}[A] = \int_{\Sigma_3} \text{Tr}\left(A \wedge dA + \frac{2}{3} A \wedge A \wedge A \right).
\end{equation}
We refer to Appendix \ref{sec:review} for more details on the 3+1 decomposition of GR and the Wheeler-DeWitt equation in both the metric and connection variables.

Let us now consider theta sectors in gravity. The main kinematical difference compared to a YM theory is that in the gravitational case the connection $A$ is complex-valued (in Lorentzian signature). However, the gauge transformations are not complexified since they must map real triads to real triads. As a result, the topological structures for the gravitational case and the YM case are identical \cite{Ashtekar:1988sw}; one must construct gravitational $\theta$-sectors to resolve quantization ambiguities
\begin{align}\label{eq:psi_CSK}
    \Psi_{\rm CSK}^\theta (A^g) = e^{i\theta w(g)}\Psi_{\rm CSK}^\theta(A)
\end{align}
analogous to Eq.~\eqref{eq:psi_YM}.

In contrast to the usual YM story, we have the explicit form of $\Psi_{\rm CSK}[A]$ in Eq.~\eqref{eq:Euclidean_CSK_with_Lambda}; the $\theta$-sector as defined by Eq.~\eqref{eq:psi_CSK} acts as a constraint. To see its impact, we evaluate the LHS of Eq.~\eqref{eq:psi_CSK}. Under a (large) gauge transformation, the Ashtekar connection transforms as 
\begin{equation}
    A \to A^g = g A g^{-1} + g d g^{-1}.
\end{equation}
Under this transformation, the Chern-Simons functional shifts by a term proportional to the winding $w(g) = n$,
\begin{equation} \label{eq:large gauge transformation}
    {\rm CS}[A^g] = {\rm CS}[A] + 8\pi^2 n, \quad n \in \mathbb{Z}.
\end{equation}
Since the CSK state \eqref{eq:Euclidean_CSK_with_Lambda} depends exponentially on \( {\rm CS}[A] \), its transformation under a large gauge transformation is given by
\begin{equation}\label{eq:quasi_periodic_psi}
    \Psi_{\rm CSK}[A^g] = e^{i\frac{12\pi^2}{\Lambda \ell_{\rm Pl}^2}n} \Psi_{\rm CSK}[A].
\end{equation}
Comparing this with our $\theta$-sector constraint in Eq.~\eqref{eq:psi_CSK}, this implies
\begin{align}
    \theta = \frac{12\pi^2}{\Lambda \ell_{\rm Pl}^2} \operatorname{mod} 2\pi.
\end{align}
Thus, we have derived Eq.~\eqref{eq:main_result}. In particular, note that the ``superselection'' choice of $\theta$ constrains the quantity $6\pi/(\Lambda \ell_{\rm Pl}^2) \in \mathbb{Z}$. In one lens, fixing $\theta$ ``quantizes'' $1/\Lambda$. Conversely, the value of $\Lambda$ determines the choice of topological $\theta$-sector. We remark that this result also resolves the ``large gauge invariance issue'' of the CSK state as previously raised by other authors \cite{Witten:2003mb, Wieland:2011de}.

The relation between $\Lambda$ and $\theta$ yields a new perspective on the cosmological constant problem. We emphasize that our statements are on the CSK state, an exact solution to the Hamiltonian constraint. Consistency of the CSK state solution with that arising from perturbative quantization requires that $\Lambda$ is robust to graviton loop corrections; if $\Lambda$ is discretely constrained by $\theta$, then there is no point in computing perturbative corrections. This is in analogy with the perturbative non-renormalization of $\theta$ in QCD\cite{Marino2015, Witten1979}. 

\section{Quantum Gravitational Hall Effect}\label{sec: quantum gravitational Hall}

In this section, we discuss potential observable consequences of the CSK state. Motivated by the presence of the Chern-Simons term, we show that the mathematics of the Hamiltonian constraint in the connection representation is reminiscent of the quantum Hall effect. The results in this section do not depend on the relation between $\Lambda$ and the $\theta$-vacua (see Eq.~\eqref{eq:main_result}) that we explored in previous sections. As we shall see, the connection between these two seemingly different physical systems motivates a new physical interpretation of the WdW equation in quantum cosmology.

From the Wheeler-DeWitt equation [Eqs. \eqref{WdW_eq} and \eqref{eq:action_of_E}], we have the relation
\begin{equation}\label{eq:WdW_condition}
    i\ell_{\rm Pl}^2 \beta \frac{\delta \Psi[A]}{\delta A^k_d} = \frac{3}{2 \Lambda} F_{ab\;k} \Psi[A].
\end{equation}
Using a semiclassical WKB approximation, one may write the wavefunction as
\begin{equation}\label{eq:WKB_approx}
    \Psi[A] \propto e^{iS_{\rm HJ}[A]},
\end{equation}
where $S_{\rm HJ}[A]$ is interpreted as a Hamilton-Jacobi functional. In that limit, we can write \eqref{eq:WdW_condition} as
\begin{equation}
    (J_{\rm H})^c_k \equiv \frac{\delta S_{\rm HJ}}{\delta A^k_d} = -\frac{3}{2\beta \Lambda \ell^2_{\rm Pl}}\epsilon^{cab}F_{ab\; k}.
\end{equation}
For the Euclidean case $\beta =-1$, this is exactly the expression for a Hall current
\begin{equation}\label{eq:Quantum_grav_Hall_current}
    (J_{\rm H})^c_k = \sigma_{\rm H}\epsilon^{cab}F_{ab\; k}
\end{equation}
with Hall conductance 
\begin{equation}\label{eq:grav_Hall_conductance}
    \sigma_{\rm H} = \frac{3}{2\Lambda \ell^2_{\rm Pl}}.
\end{equation}
Furthermore, if $\theta$ is fixed, the gravitational Hall conductance is \emph{quantized} via Eq.~\eqref{eq:main_result}.

For the Euclidean case, the CSK state \eqref{eq:Euclidean_CSK_with_Lambda} has the form \eqref{eq:WKB_approx} with $S_{\rm HJ}[A] = 3{\rm CS}[A]/(2\Lambda \ell^{2}_{\rm Pl})$ without any approximation. In other words, the Chern-Simons action for the self-dual connection is precisely the Hamilton-Jacobi functional of general relativity \cite{Smolin:2002sz}. So, for $\Psi_{\rm CSK}[A]$, the previous calculations are exact and the physics of the Hamiltonian constraint is analogous to the Hall effect. 

In standard Hall-effect physics, the Hall current is the (transverse) response to an externally applied electromagnetic field to the conductor. In the gravitational case, one can think of $A_a^i$ as an ``electromagnetic potential'' field that is applied to constant time hypersurfaces. This field gives rise to the \emph{quantum gravitational Hall current} \eqref{eq:Quantum_grav_Hall_current}, where the current strength is controlled by the curvature of the connection. This supports the classical picture of GR, for the ``response'' to the cosmological constant is the induced curvature of the spacetime.

We now show that the quantum gravitational Hall current\eqref{eq:Quantum_grav_Hall_current} is proportional to the probability current associated with $\Psi_{\rm CSK}[A]$ in the Euclidean case. In the usual metric variables, one can define the superspace (or probability) current \cite{DeWitt}
\begin{equation}\label{eq:KG_current_canonical}
    J_{ab}= \frac{i}{2}G_{abcd}\left(\Psi^*[h_{ab}]\frac{\delta \Psi[h_{ab}]}{\delta h_{cd}} - \frac{\delta \Psi^*[h_{ab}]}{\delta h_{cd}}\Psi[h_{ab}]\right),
\end{equation}
where $G_{abcd}$ is the DeWitt metric. This current is conserved on solutions of the WdW equation. In the minisuperspace, \eqref{eq:KG_current_canonical} reduces to
\begin{equation}\label{eq:KG_current_minisuperspace}
    j_A= \frac{i}{2}\left(\Psi^*[a^2,\phi]\nabla_{A}\Psi[a^2, \phi] -  \Psi^*[a^2, \phi]\nabla_A\Psi[a^2, \phi]\right),
\end{equation}
where $\nabla_A$ stands for derivatives with respect to the scale factor squared $a^2$ and any other possible fields, collectively denoted by $\phi$. Eq.~\eqref{eq:KG_current_minisuperspace} can be used to define the boundary conditions in minisuperspace quantum cosmology, e.g. the Vilenkin wavefunction is such that the flux $j_A$ points outward at singular boundaries (only outgoing modes) \cite{Vilenkin:1987kf}.

Consider now 
\begin{equation}\label{eq:prob_current_connectionrep}
    j^a_{i}[\Psi] = \frac{i}{2}\left(\Psi^*[A]\frac{\delta}{\delta A^i_a}\Psi[A] - \Psi[A]\frac{\delta}{\delta A^i_a}\Psi^*[A]\right),
\end{equation}
which is the equivalent of \eqref{eq:KG_current_canonical} in terms of the self-dual connection.
Evaluating this current in the (normalized) Euclidean CSK state gives
\begin{equation}\label{eq:KG_current_selfdual}
    j^a_i[\Psi_{\rm CSK}] = - \frac{3}{2\Lambda \ell^2_{\rm Pl} }\epsilon^{abc}F_{ab,\;i}. 
\end{equation}
Comparing with \eqref{eq:Quantum_grav_Hall_current} we find that $j_i^a$ is proportional to the gravitational Hall current,
\begin{equation}\label{eq:KG_Hall_current_relation}
    j_i^a[\Psi_{CSK}] = - (J_{\rm H})^a_i.
\end{equation}
Note that \eqref{eq:KG_current_selfdual} vanishes for the Lorentzian signature CSK state, while $J_{H}$ in \eqref{eq:Quantum_grav_Hall_current} is complex valued in that case. 

\section{Discussion}

In this work, we demonstrated that the topological $\theta$ sectors and the cosmological constant are intimately linked via Eq.~\eqref{eq:main_result}.  This suggests that perturbative graviton loop corrections to the cosmological constant do not renormalize $\Lambda$ and could potentially solve the perturbative UV part of the cosmological constant problem. We will pursue this issue further in future work.

We also found a correspondence between the quantum Hall effect and the Hamiltonian constraint, with the probability current of the Euclidean Chern-Simons-Kodama state playing the role of a gravitational Hall current and $3/(2\Lambda \ell_{\rm Pl}^2)$ the Hall conductance. Furthermore, for fixed $\theta$, Eq.~\eqref{eq:main_result} implies that the gravitational Hall conductance is quantized. This again suggests the robustness of $\Lambda$ against corrections, similar to the topological protection of the Hall conductance in the quantum Hall effect.

The presence of topological $\theta$-terms also suggest the possibility of interesting topological observables, such as those in 3D topological insulators \cite{Sekine2021,Qi2008,Burkov2015,Wen1995,Zee1995}. If $\theta$ is allowed to vary (like an axion), this can give rise to topological currents. For instance, finite regions distinguished by distinct values of $\theta$ will support topological currents on the boundary. We leave the discussion of these effects and their potential relation to the cosmological constant to future work.

For generic $\theta \neq 0$, the $\theta$-vacuum breaks parity, similar to what happens in QCD. However, CP is preserved for $\theta = 0$ and $\theta = \pi$ (see \cite{Soo:1995ga} for discrete transformations of the gravitational action in self-dual variables). Therefore, we can construct a ``non-trivial'' CP-invariant vacuum state using our proposal in Eq.~\eqref{eq:main_result}
\begin{equation}
    \theta = \pi \implies \Lambda = \frac{12\pi}{\ell_{\rm Pl}^2 (1+2n)}.
\end{equation}
for $n \in \mathbb{Z}$. This is in addition to the trivial $\theta = 0$ case \cite{Wieland:2011de}. 

Throughout this work, we discuss $\Psi[A]$ in the $A$ representation. We remark that it is difficult, if not impossible, to see the connection between $\Lambda$ and topology by looking at wavefunctions $\Psi(E)$ in the $E$ representation. As discussed in Ref.~\cite{Ashtekar:1988sw}, the configuration space in the $E$ representation is disconnected. There are no solutions connecting triads associated with connections of different homotopy classes. This is the reason why the bare cosmological constant $\Lambda$ seems unconstrained and unrelated to topology in the usual metric formulation of GR.

According to the no-boundary proposal \cite{Hartle:1983ai}, the Hartle-Hawking wavefunction is obtained from an Euclidean path integral over metrics with appropriate boundary conditions. This is easily carried out in the minisuperspace approach, where the sum is over scale factors. An analogous computation of the CSK state and hence derivation of the relation with the theta sector would be challenging, since the relevant configuration variable we considered is the connection, not the metric. The wavefunction in the metric representation can only be recovered after integrating over connections. In fact, both the Hartle-Hawking and Vilenkin wavefunctions in minisuperspace cosmology \cite{Hartle:1983ai, Vilenkin:1987kf,Vilenkin:1994rn} can be obtained from the CSK state in such a way, as shown in \cite{Magueijo:2020ugp}.

\section{Acknowledgments}

We dedicate this work to the memory of James Simons who encouraged one of us (SA) to pursue this work. We thank Abhay Ashtekar, Saurya Das, Keshav Dasgupta, Laurent Friedel, Antal Jevicki, Jo\~ao Magueijo, Horatiu Nastase, Philip Phillips, and David Spergel for discussion and comments on a previous version of this work. SA and HB were supported by the Simons Foundation through Award No. 896696. 

\bibliography{biblio}

\appendix

\section{The Gravitational Hamiltonian with a Cosmological Constant}\label{sec:review}

To make the discussion in this letter self-contained, we review the equivalence between the Arnowitt-Deser-Misner (ADM) Hamiltonian formalism \cite{ADM:1959} and that of Ashtekar \cite{Ashtekar:1986, Ashtekar:1987} (see also \cite{Sen:1982qb}). The Ashtekar formulation of general relativity recasts the Hamiltonian (i.e. the canonical) formulation in terms of a connection variable \( A^i_a \) and its conjugate momentum, the densitized triad \( \tilde{E}^a_i \), instead of the conventional metric variables of the ADM formalism (see \cite{Ashtekar:1991hf,Kiefer:2004xyv,Rovelli:2004tv, Thiemann:2007pyv} for references on canonical quantum gravity). We emphasize that the Ashtekar formulation is non-perturbative and allows us to sidestep laborious computations in the typical perturbative approaches based on expansions around flat spacetime. 

In the ADM formalism, spacetime is foliated by a family of non-intersecting space-like hypersufaces from which a congruence of curves generated by $t^\alpha$ emanates. Since the congruence may not be orthogonal to the hypersurfaces, we can write $t^\mu = N n^\mu + N^a e^\mu_a$, where $n^\mu$ is the unit normal to the hypersurfaces, $N^a$ is the shift vector, $N$ is the lapse function, and $e^\mu_a = \partial x^\mu/\partial y^a$ where $y^a$ are the coordinates on the hypersurfaces (the index $a$ runs from 1 to 3). The spacetime metric splits as
\begin{equation}
    ds^2 = -N^2 dt^2 + q_{ab}\left(N^a dt + dy^a\right)\left(N^b dt + dy^b\right)
\end{equation}
where $t$ parametrizes the congruence of curves generated by $t^\mu$, i.e. $t^\mu \partial_\mu t=1$. In terms of 3-dimensional quantities, the Einstein-Hilbert action becomes 
\begin{align}
    S_{\rm EH} &= \frac{1}{16\pi G} \int d^4x \sqrt{-g} \; R \nonumber\\
    &= \frac{1}{16\pi G} \int dt d^3y \;N\sqrt{q} \left(R^{(3)} + K^{ab}K_{ab}K^{ab}- K^2\right),
\end{align}
where $R^{(3)}$ is the spatial Ricci scalar, $K_{ab} = \nabla_b^{(3)}n_b$ the extrinsic curvature of the hypersurfaces of constant $t$, and $K = h^{ab}K_{ab}$. The spatial covariant derivative $\nabla^{(3)}$ is defined from the three-dimensional Christoffel symbols associated with $h_{ab}$. For the sake of presentation we will consider the vanishing cosmological constant case first.

In the Hamiltonian formalism for GR, the Lie derivative of the 3-metric with respect to $t$ plays the role of a time derivative, so we define $\dot{q}_{ab} = \mathcal{L}_t q_{ab}$. The canonical momentum associated with it is
\begin{equation}
    \pi^{ab} = \frac{1}{16\pi G} \sqrt{q}\left(K^{ab} - K h^{ab}\right).
\end{equation}
Using this, we can compute the so-called ADM Hamiltonian which is given by
\begin{equation}
H_{\text{ADM}} = \int d^3x \left(N \mathcal{H} + N^a \mathcal{H}_a \right),
\end{equation}
where \( \mathcal{H} \approx 0\) is the scalar (or Hamiltonian) constraint and \( \mathcal{H}_a \approx 0\) is the diffeomorphism constraint. Explicitly,
\begin{align}
\mathcal{H} &= \frac{16\pi G}{\sqrt{q}} \left( q_{ac} q_{bd} - \frac{1}{2} q_{ab} q_{cd} \right) \pi^{ab} \pi^{cd} - \frac{\sqrt{q}}{16\pi G} R^{(3)}, \\
\mathcal{H}_a &= -2 \nabla_b^{(3)} \pi^b_a.
\end{align}

In the Ashtekar formalism, the dynamical variables are \( A^i_a \), an SU\((2) \) connection, and its conjugate momentum \( \tilde{E}^a_i = \sqrt{q}E^a_i \), related to the spatial metric via
\begin{equation}
q^{ab} = E^a_i E^b_j \delta^{ij}.
\end{equation}
The tensor density $\tilde{E}^a_i$ is often called the (densitized) triad. The Ashtekar connection \(A^i_a\) is obtained from a canonical transformation of the phase space in terms of ADM variables in general relativity \cite{Ashtekar:1986,Immirzi:1996di}:
\begin{align}
A^i_a = \Gamma^i_a -\frac{1}{\beta}K^i_a,
\end{align}
where $\Gamma^i_a = -(1/2)\epsilon^{i\;\;k}_{\;j}\Gamma^j_{lk}E_a^l$ is the dual of the spin connection determined by the triad, and \(K^i_a = K_{a}^bE_{b}^i\) is the extrinsic curvature expressed in the \(\text{SU}(2)\) internal frame. We introduced the constant $\beta$ to handle both the Lorentzian $\beta= i$ and the Euclidean ($\beta =-1$) cases\footnote{We assume that $iS_{\rm EH}^{\rm Lorentzian} = -S_{\rm EH}^{\rm Euclidean}$, so the Euclidean Einstein-Hilbert Lagrangian differs from the Wick-rotated Lorentzian one by an overall sign. The constant $\beta$ is related to the so-called Barbero-Immirzi parameter \cite{BarberoG:1994eia, Immirzi:1996di,Immirzi:1996dr}.}. The Hamiltonian in this formulation takes the form
\begin{equation}
H_{\text{Ashtekar}} = \int d^3x \left( N \tilde{\mathcal{H}} + N^a \mathcal{H}_a + \lambda^i \mathcal{G}^i \right),
\end{equation}
where \( \mathcal{G}^i\approx0 \) is the Gauss constraint enforcing SU\((2) \) gauge invariance, where
\begin{equation}
\mathcal{G}^i = \mathcal{D}_a \tilde{E}^a_i.
\end{equation}
The Hamiltonian constraint in this formulation is given by
\begin{equation}
\tilde{\mathcal{H}} = \frac{\beta^2}{16\pi G\sqrt{|\Tilde{E}|}} \epsilon^{ijk} \tilde{E}^a_i \tilde{E}^b_j F_{ab}^k,
\end{equation}
where \( F_{ab}^k \) is the curvature of the Ashtekar connection.

To recover the Lorentzian metric signature, one needs to impose reality conditions on $A^i_a$ \cite{Ashtekar:1987}. Then, expressing \( F_{ab}^k \) in terms of the Ricci curvature, one can show that the Ashtekar Hamiltonian constraint reproduces the ADM constraint structure. Thus, the two Hamiltonians describe the same classical theory, albeit in different phase-space representations \cite{Ashtekar:1987}. The Ashtekar variables simplify the constraint equations of general relativity, providing a gauge-theoretic perspective that retains full equivalence with the ADM formulation while offering a more compact and efficient framework for canonical quantization. 

The canonical transformation preserves the symplectic structure of the phase space, ensuring equivalence with the original formulation. Specifically, the symplectic form in the ADM variables can be rewritten with Ashtekar variables as
\begin{equation}
\int \pi^{ab} \delta q_{ab} \sim \int \tilde{E}^a_i \delta A^i_a.
\end{equation}
The new canonical variables satisfy the fundamental Poisson bracket
\begin{equation}\label{Poisson_bracket_new_variables}
\{A^i_a(x), \tilde{E}^b_j(y)\} = 8\pi G\beta \delta^b_a \delta^i_j \delta^{(3)}(x-y),
\end{equation}
demonstrating that the transformation does not alter the underlying Hamiltonian structure. Recall that $\beta = i$ ($\beta = -1$) in the Lorentzian (Euclidean) case.

The ``dynamics'' of the theory are governed by the constraints, with the Hamiltonian constraint \(\mathcal{\tilde{H}} \approx 0\) playing a central role. In the presence of a cosmological constant \(\Lambda\), the Hamiltonian constraint becomes \cite{Thiemann:2007pyv}
\begin{equation}
\mathcal{\tilde{H}} = \frac{\beta^2}{16\pi G \sqrt{|\tilde{E}}|}\epsilon^{ijk} \tilde{E}^a_i \tilde{E}^b_j \left(F_{ab}^k + \frac{\Lambda}{3} \epsilon_{abc} \tilde{E}^c_k \right).
\end{equation}

Upon quantization, we can seek a quantum solution to the Wheeler-DeWitt equation 
\begin{equation}\label{WdW_eq}
\hat{\tilde{\mathcal{H}}} \Psi[A] = 0.
\end{equation}
From \eqref{Poisson_bracket_new_variables}, we have that in the connection representation,
\begin{equation}\label{eq:action_of_E}
    \hat{\tilde{E}}^b_j(y) = -i8\pi G \hbar \beta\frac{\delta}{\delta A^j_b(y)}.
\end{equation}
With this action on $\Psi[A]$, assuming non-degenerated triads, \eqref{WdW_eq} implies that
\begin{equation}
    \left(F_{ab\;k} - i\frac{\Lambda 8\pi G \hbar\beta}{3}\epsilon_{abc}\frac{\delta}{\delta A_c^k}\right)\Psi[A] = 0.
\end{equation}
So, a solution to \eqref{WdW_eq} is given by
\begin{equation}\label{kodama_wavefunction}
    \Psi[A] = \mathcal{N} e^{-\frac{3i}{2\beta\Lambda \ell^2_{\rm Pl}}\rm{CS}[A]},
\end{equation}
where $\mathcal{N}$ is a normalization constant, $\ell_{\rm Pl} = \sqrt{8\pi G \hbar}$ is the reduced Planck length, and ${\rm CS}[A]$ is the Chern-Simons functional of $A_a^i$,
\begin{equation}
    {\rm CS}[A] = \int_{\Sigma_3} \text{Tr}\left(A \wedge dA + \frac{2}{3} A \wedge A \wedge A \right).
\end{equation}
where $\Sigma_3$ is the 3D manifold of the 3+1 decomposition of spacetime.

The wavefunction \eqref{kodama_wavefunction} is called the Chern-Simons-Kodama (CSK) \cite{Kodama}. It is an \emph{exact} state that solves all the GR constraints in a non-perturbative way. In this letter, special attention is given to the Euclidean case ($\beta = -1$), 
\begin{equation}\label{eq:App_Euclidean_CSK_with_Lambda}
    \Psi_{\rm CSK}[A] = \mathcal{N} e^{\frac{3i}{2\Lambda \ell^2_{\rm Pl}}{\rm CS}[A]}.
\end{equation}

\end{document}